\documentclass[aps,pra,twocolumn,eqsecnum,superscriptaddress]{revtex4-2}
\usepackage{float}
\usepackage{amsmath}
\usepackage{amssymb}
\usepackage{graphicx}
\usepackage{mathrsfs}
\usepackage{color}
\usepackage{hyperref}
\usepackage{lipsum}
\usepackage[procnames]{listings}
\usepackage{layouts}
\usepackage{epstopdf}
\usepackage{dsfont}
\usepackage{bbm}
\usepackage{ulem}
\usepackage{comment}
\usepackage{tabularx}

\newcommand{\leqnomode}{\tagsleft@true\let\veqno\@@leqno}%
\newcommand{\reqnomode}{\tagsleft@false\let\veqno\@@eqno}%
\newcommand*{\compress}{\@minipagetrue}
\makeatother

\newcommand{\bra}[1]{{\left\langle{#1}\right\vert}}
\newcommand{\ket}[1]{{\left\vert{#1}\right\rangle}}

\DeclareUnicodeCharacter{2009}{\,} 
\newcommand{\qu}[1]{``#1''}
\newcommand{\Tr}{\text{Tr}}

\newcolumntype{M}{>{\displaystyle}r@{\hspace{1pt}\pm\hspace{1pt}}l}

\begin{document}

\title{Hierarchical Verification of Non-Gaussian Coherence in Bosonic Quantum States}

\author{Beate E. Asenbeck} 
\affiliation{Laboratoire Kastler Brossel, Sorbonne Universit\'e, CNRS, ENS-Universit\'e PSL, Coll\`ege de France, 4 Place Jussieu, 75005 Paris, France}
\author{Luk\'a\v s Lachman}
\affiliation{Laboratoire Kastler Brossel, Sorbonne Universit\'e, CNRS, ENS-Universit\'e PSL, Coll\`ege de France, 4 Place Jussieu, 75005 Paris, France}
\affiliation{Department of Optics, Palack\'y University, 17. Listopadu 12, 771 46 Olomouc, Czech Republic}
\author{Ambroise Boyer}
\affiliation{Laboratoire Kastler Brossel, Sorbonne Universit\'e, CNRS, ENS-Universit\'e PSL, Coll\`ege de France, 4 Place Jussieu, 75005 Paris, France}
\author{Priyanka Giri}
\affiliation{Laboratoire Kastler Brossel, Sorbonne Universit\'e, CNRS, ENS-Universit\'e PSL, Coll\`ege de France, 4 Place Jussieu, 75005 Paris, France}
\author{Alban Urvoy}
\affiliation{Laboratoire Kastler Brossel, Sorbonne Universit\'e, CNRS, ENS-Universit\'e PSL, Coll\`ege de France, 4 Place Jussieu, 75005 Paris, France}
\author{Radim Filip}
\affiliation{Department of Optics, Palack\'y University, 17. Listopadu 12, 771 46 Olomouc, Czech Republic}
\author{Julien Laurat}
\email[email: ]{julien.laurat@sorbonne-universite.fr}
\affiliation{Laboratoire Kastler Brossel, Sorbonne Universit\'e, CNRS, ENS-Universit\'e PSL, Coll\`ege de France, 4 Place Jussieu, 75005 Paris, France}

\begin{abstract}
Non-Gaussianity, a distinctive characteristic of bosonic quantum states, is pivotal in advancing quantum networks, fault-tolerant quantum computing, and high-precision metrology. Verifying the quantum nature of a state, particularly its non-Gaussian features, is essential for ensuring the reliability and performance of these technologies. However, the specific properties required for each application demand tailored validation thresholds. Here, we introduce a hierarchical framework comprising absolute, relative, and qubit-specific thresholds to assess the non-Gaussianity of local coherences. We illustrate this framework using heralded optical non-Gaussian states with the highest purities available in optical platforms. This comprehensive framework presents the first detailed evaluation of number state coherences and can be extended to a wide range of bosonic states.
\end{abstract}

\maketitle

\par  \textit{Motivation.---}The rapid advancements in quantum information science hold the promise of surpassing classical systems in computational speed, security, and precision. Among the fundamental properties of quantum states, non-Gaussianity \cite{Lachman2022,Walschaers2021} is a key indicator of non-classical behavior. It reflects the deviation from Gaussian states, such as coherent and squeezed states, which can be efficiently simulated classically and are insufficient for achieving quantum advantage in many tasks \cite{Cerf2009,Chabaud2021,Baragiola2019,GarciaAlvarez2020}. Various platforms, including ions \cite{Rojkov2024, Harty2014, Fluehmann2019}, neutral atoms \cite{Magro2023}, superconducting qubits \cite{Wang2017, CampagneIbarcq2020, Reglade2024}, mechanical systems \cite{Chu2018}, and photons \cite{Morinreview, Ourjoumtsev2006, Huang2015,Kawasaki2022,Cotte2022,Takase2022}, have demonstrated the ability to generate non-Gaussian states, showcasing the remarkable progress achieved in quantum state engineering.

In this context, characterizing non-Gaussian states has attracted considerable interest. Unlike Gaussian states, which have a well-defined canonical representation through the covariance matrix, non-Gaussian states require more sophisticated methods for their classification and quantification \cite{Lvovsky2020,Walschaers2021}. A widely used approach is the analysis of the Wigner function negativity, where negative regions indicate non-classical features \cite{Genoni2013}. The negative volume of the Wigner function can provide a global quantitative measure of non-Gaussianity \cite{Zaw2024}. Another tool is the Hilbert-Schmidt distance or quantum relative entropy, which measures the dissimilarity between a quantum state and a reference Gaussian state \cite{Genoni2010}. 

Recently, a stellar hierarchy was introduced \cite{Chabaud2020} as a structured framework for classifying non-Gaussianity by analyzing the distribution of zeros in the Husimi function. This method identifies successive levels of non-Gaussian features, offering a comprehensive means of comparison. However, while this hierarchy is effective for assessing global non-Gaussianity and informs on the number of single-excitation additions needed for their engineering, it does not address whether a given state is well-suited for specific applications. For instance, binomial states, which are specific superpositions of Fock states, are non-Gaussian and prime candidates for quantum error correction \cite{Michael2016, Zhuang2018}. However, non-Gaussianity alone is insufficient to determine their suitability for this purpose. A four-photon state shares the same non-Gaussianity level in the stellar hierarchy as the simplest logical binomial state, yet only the latter is applicable for error correction. This highlights the need for a hierarchical framework that evaluates both non-Gaussianity and the specific properties required for various quantum applications.

\begin{figure}[b!]
\includegraphics[width=0.92\columnwidth]{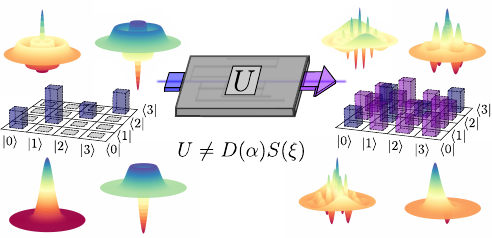}
\caption{Non-Gaussian coherence creation. A Wigner-symmetric state, initially characterized by a diagonal density matrix in the Fock basis, undergoes a non-Gaussian unitary transformation $U$ that cannot be written as a combination of the Gaussian operations of displacement $D$ and squeezing $S$. The resulting state exhibits non-Gaussian coherences, which can be verified via the criteria developed in this work.}\label{fig1} 
\end{figure}	

A crucial property in this context is quantum coherence, represented by the off-diagonal elements of a quantum state in a chosen basis. Figure \ref{fig1} illustrates the difference between non-Gaussian states that are symmetric in phase space, which lack coherences, and the states of interest in this work, which exhibit superpositions in the Fock basis. Coherences are fundamental to many quantum tasks as they underpin the creation of superpositions, which are essential for processes such as error correction and sensing \cite{Aberg2006,Giovannetti2006,McCormick2019, Michael2016}. Here, we propose and experimentally test a hierarchical framework for assessing non-Gaussian coherences in superpositions of Fock states.

\par  \textit{Principle and experimental state creation.---}To classify and evaluate the coherence of non-Gaussian states, we draw from the principles of quantum resource theories, which categorize states and operations into \textit{free} or \textit{resourceful} based on their accessibility within a given physical system \cite{HORODECKI2012,Streltsov2017}. Free states, which can be generated using a restricted set of operations, act as the baseline for comparison. In contrast, resourceful states requires more complex operations and they offer unique functionalities that cannot be achieved using free operations and states alone. In our study, we focus on coherences in the Fock basis as the key resource for distinguishing the states. 

More specifically, examples are illustrated in Fig. \ref{fig1}. The bottom-right state represents a superposition between the vacuum and single-photon states, which cannot be generated by applying Gaussian operations, namely squeezing and displacement, to the vacuum and single-photon states shown on the bottom left. By progressively broadening the definition of free states, we construct a hierarchy, enabling finer distinctions among the levels of non-Gaussian coherences. Designating Gaussian operations as free aligns with standard definitions of non-Gaussianity, while extending free states beyond the vacuum introduces deviations that refine our understanding of coherence in specific contexts. This distinction allows us to classify highly non-Gaussian Fock states as insufficient for certain tasks where coherence is at the core, emphasizing the importance of application-specific metrics over global measures. 

To test this framework, we use experimentally generated heralded non-Gaussian states from optical parametric oscillators (OPOs) operating below threshold in combination with single-photon detectors \cite{Huang2015,Zapletal2021,Huang2016}. These states represent some of the most non-Gaussian optical states achievable today, characterized by very low vacuum-admixtures and single-photon purities exceeding 90\%. Nonetheless, achieving high coherences in these systems remains a significant challenge compared to other systems  \cite{Srinivas2021,Magro2023,Fluehmann2020,Wang2017}. Utilizing our optical platform, we produced superposition states $\propto \ket{0}-\ket{1}$  \cite{Darras22} and $\propto \ket{0}+\ket{2}$ \cite{Huang2015} through two distinct methods of coherence generation involving displacement operations and $N$-photon heralding \cite{supp}. Rigourous analysis of the non-Gaussian features in these states is made possible by direct homodyne tomography, which enables precise density matrix reconstruction with minimal errors \cite{supp}.

\vspace{0.12 cm} 
\par  \textit{Absolute non-Gaussian coherence criteria.---}To establish non-Gaussian coherence criteria, we need to define free states and operations, and a specific coherence measure. Coherences can be either distributed over the whole Hilbert space of a system, or can be localized as interference terms between two Fock-state probabilities. Although both types of coherences are relevant, we focus on the localized case, which translates to target states of the form $(1/\sqrt{2})(\ket{n_1}+e^{i\phi}\ket{n_2})$, representing the maximal local coherence between two Fock states.

The easiest choice of free states consists of states that possess no coherences, meaning all Fock states, with an excitation cut-off at $N_{\text{max}}$. Arbitrary free operations can be applied to these Fock states $\ket{f}$ and their linear combinations. Depending on the allowed operations, we can define two sets of free states: allowing both squeezing $S(\xi)$ and displacement $D(\alpha)$ results in the Gaussian free states, $F^G$, while excluding squeezing yields the classical set, $F^C$.

We define the rank $l$ in the hierarchy as the superposition length of the target state $n_2-n_1$, such that $l \in [1, n_2-n_1]$, which we call the $L$-hierarchy \cite{supp}. A detailed derivation and extended discussions are provided in the companion paper \cite{Lukas2024}. This hierarchy defines all states $\ket{\psi_l}$ of Eq. \ref{LTargetStates} as equal in rank. To create a proper hierarchy, each lower ranked state  $\ket{\widetilde{\psi}_l}$ has to be included into the set of free states for the next rank, leading to the definition of the set of Gaussian free states $F^{G}_l$ of rank $l$

\begin{eqnarray}
&\ket{\psi_l} \propto \ket{n}+\ket{n+l} \quad n \in [0,N_{\text{max}}-l] \label{LTargetStates}\\
 &\ket{\xi,\alpha,\widetilde{\psi}_l}  = S(\xi)D(\alpha)\ket{\widetilde{\psi}_l} \in F_l^G, \label{LmaxFreeStates} \\
 &\text{with} \quad \ket{\widetilde{\psi}_l} = \sum_{f=i}^{i+l-1}c_f \ket{f} \quad i \in [0,N_{\text{max}}-l] .  \nonumber
\end{eqnarray}

For a rank $l$, superpositions of a length of maximally $l-1$ are defined as free. We note here that those free-state superpositions do not have to be local. A similar approach applies to define the set of classical free states $F^C_l$.

\begin{figure}[b!]
\includegraphics[width=0.95\columnwidth]{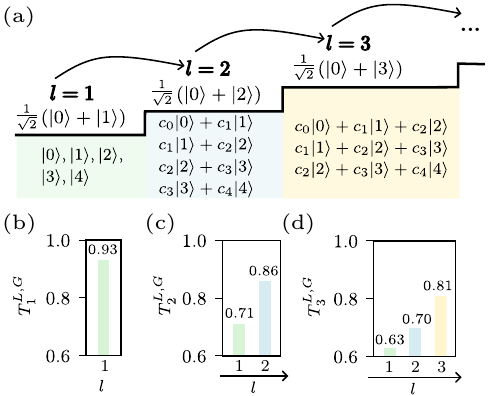}
\caption{Hierarchy of non-Gaussian coherence criteria. For superpositions of the form $\ket{0}+\ket{n_2}$, the hierarchy is determined by the distance $l=n_2$ between the two Fock excitations. (a) The hierarchy is visualized with ideal target states displayed above and basis states for $N_{\textrm{max}}=4$ shown below the staircase. (b)-(d) Threshold values for each target state are obtained by optimizing over the set of free operations. The hierarchy ranks are indicated using corresponding colors.}\label{fig2} 
\end{figure}

\begin{figure*}[htpb!]
\includegraphics[width=1.7\columnwidth]{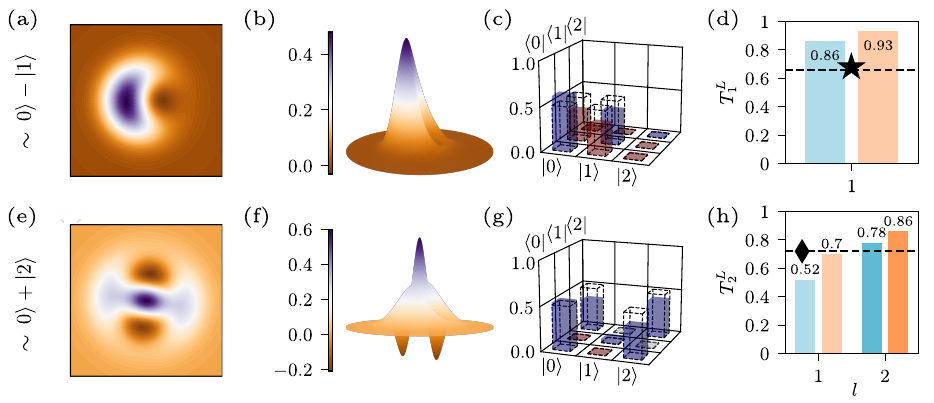}
\caption{Experimental data and absolute non-classical and non-Gaussian coherence threshold. (a) and (b) display the Wigner function of an experimentally generated superposition state $\propto \ket{0}-\ket{1}$ from the top view and in 3D, respectively. Similarly, (e) and (f) show the Wigner function for the state $\propto \ket{0}+\ket{2}$. (c) and (g) provide the corresponding absolute values of the density matrices, with blue indicating positive values and red negative ones. The ideal state is represented in dotted bars. (d) and (h) illustrate the hierarchical certification of coherence based on the rejection of classical or Gaussian dynamics in the measurement of the coherence $C_{0n}$ with $n\in \left\{1,2\right\}$. Blue bars corresponds to the non-classical thresholds, and orange bars to the non-Gaussian thresholds. The star and diamond are the experimental results. Error bars are smaller than the data points, as discussed in the supplementary material \cite{supp}.}\label{fig3} 
\end{figure*} 

Next, a measure of coherence has to be defined in the context of this hierarchy. This measure should be maximized by the target states of the rank. To assess if a state $\rho'$ achieves a certain rank, the best value achieved for free states for this rank has to be computed, thereby defining the threshold. For a measure of local coherences, we can define the operator $X_{m,n}(\phi)\equiv |m\rangle \langle n|\exp(i\phi)+|n\rangle\langle m|\exp(-i\phi)$, which corresponds to a projective measurement along the equator of a qubit on a Bloch sphere with poles $\ket{n_1}$ and $\ket{n_2}$. The local coherence measure $\mathcal{C}$ of any state $\rho$ can then be defined as applying $X$ on the state and identifying the highest coherences around the Bloch sphere equator, \textit{i.e.}, $\theta =\pi/2$, as
\begin{eqnarray}
\mathcal{C}_{n_1,n_2}(\rho)=\frac{1}{2}(&\max_{\phi}\Tr[X_{n_1,n_2}(\phi) \; \rho]\nonumber \\
-&\min_{\phi}\Tr[X_{n_1,n_2}(\phi) \; \rho]). \label{Cdef}
\end{eqnarray}
This can be interpreted as an interferometric measurement with a phase scan $\phi$, with $\mathcal{C} \in [0,1]$ the contrast of the fringes. The target state achieves the ideal value of unity. The threshold $T_{l}^{L,G}$ ($T_{l}^{L,C}$) that any state $\rho'$ must surpass to achieve rank $l$ has now also to be maximized over the initial Fock state $m$ of the free states defined by Eq. \ref{LmaxFreeStates}, including maximizing over the superposition weights $c_j$, the number $m$, the squeezing $\xi$ and the displacement $\alpha$, as well as the statistical mixing of those free states \cite{supp}.

As an example in Fig. \ref{fig2}, we set $N_{\text{max}}=4$ and evaluate the hierarchy up to the rank $l=3$. As shown in Fig. \ref{fig2} (b)-(d), the maximal threshold, \textit{i.e.}, the threshold of the maximal rank, consistently decreases for target states with higher $n_2$. This trend has been numerically verified for $n_1=0$, $n_1 \neq 0$, and $n_2$ up to 7. This result leads to the somehow counterintuitive fact that  thresholds are higher for short superposition lengths $l$. This decrease reflects the importance of higher and longer superpositions. It shows that they become progressively more difficult to produce using Gaussian operations. Displacement and squeezing can generate a coherent superposition of vacuum and single-photon states with high fidelity, effectively raising the threshold for verifying non-Gaussian coherence. However, displacement and squeezing also produce unwanted excitations when creating coherence between higher-photon states, reducing the maximum achievable coherence and threshold.

We now test the thresholds on our experimental states, as shown in Fig. \ref{fig3}. The first row (a)-(c) shows the state proportional to $\ket{0} - \ket{1}$, which achieves a fidelity of 82\% with the ideal superposition. The second row (e)-(g) depicts the state proportional to $\ket{0} + \ket{2}$, with a fidelity of 83\%. Both states exhibit limited vacuum admixture and these fidelities show that the setup produces state-of-the-art optical states. Figure \ref{fig3}(d) provides the absolute non-classical (in blue) and non-Gaussian (in orange) coherence thresholds, along with the experimental value $\mathcal{C}_{0,1}^{\text{exp}}=0.66$. Despite the high purity of the state, neither threshold is beaten. Similarly, Fig. \ref{fig3}(h) provides the thresholds for both ranks $l=1$ and $l=2$, alongside the coherence value $\mathcal{C}_{0,2}^{\text{exp}}=0.72$. For this state, the absolute non-Gaussian threshold of the first rank $l=1$ is surpassed, but the second rank $l=2$ is not reached due to residual unwanted Fock components and phase noise.

The absolute criterion presented so far is highly demanding. To provide intermediate steps, we introduce two additional \textit{relative} criteria. These leverage additional accessible information from the measured density matrix.  For these criteria, we will only use the strongest hierarchy (highest possible $l$), allowing for the largest set of free states to compute the thresholds. 

\vspace{0.1 cm} 
\par \textit{Relative non-Gaussian coherence criteria.---}Rather than relying exclusively on the operator $O$, we now take into account additional properties of the experimental states. For example, we may impose the condition that the free state must have a specific amount of two-photon component, matching what is measured in the experimental state. This additional condition effectively narrows the set of free states, making it more difficult for states to reach high values of the criterion. Each condition introduces a new dimension to the criterion, transforming the original absolute criterion into a relative one. 

The new $d$-dimensional measurement $P_c$ is written as a convex linear combination of all measured properties, such that $P_c^{\{g_i\}}{=}\sum_i g_i P_i'$, where $g_i$ are the weights in the convex sum, and $P_i'$ the individual measurements. Any threshold derived from this convex sum represents a relative (non-classical) non-Gaussian coherence criterion, conditioned on $i \neq 1$ properties. Note that the weights $g_i$ are optimized for each experimental input state, such that the non-Gaussian threshold can be computed as:
\begin{eqnarray}
T_{l}^{L,G} = \max_{\ket{F}^{L,G}_{m,l} \in F^{G}_l} P_c^{\{g_i\}}(\ket{F}^{L,G}_{m,l}).
\end{eqnarray}
The same procedure is applied to determine the non-classical threshold $T_{l}^{L,C}$.

Specifically for our states, we incorporate two types of additional measurements. The first is the Fock state projection \(P_m\), which represents the probability of the state being in a specific Fock state $\ket{m}$. The second is a cumulative projection \(P_{n+}\), which accounts for the combined probabilities of the state being in any Fock states from a particular excitation level \(n\) \cite{supp}. 

\begin{figure}[t!]
\includegraphics[width=0.95\columnwidth]{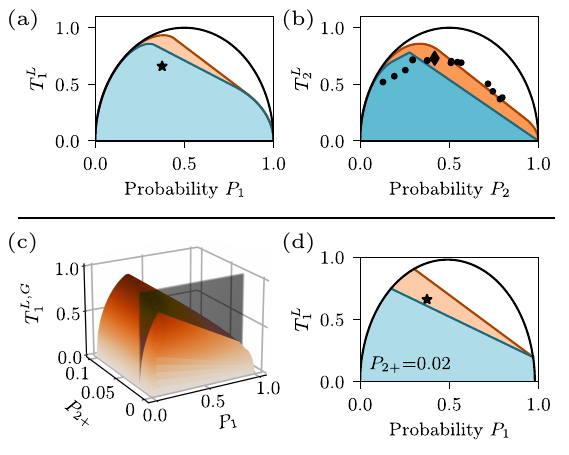}
\caption{Multi-dimensional relative thresholds. (a) and (b) correspond to two-dimensional criteria, with non-classical (blue) and non-Gaussian (orange) thresholds. In (a), the experimental state corresponds to the state $\propto \ket{0}-\ket{1}$ while (b) shows a set of experimental states $\propto \ket{0}+\ket{2}$, with diamond corresponding to the states in Fig. \ref{fig3}. $P_1$ and $P_2$ are the single- and two-photon probabilities, respectively. (c) and (d) provide three-dimensional (non-classical) non-Gaussian thresholds for the state $\propto \ket{0}-\ket{1}$. In (c), the threshold $T_1^{L,G}$ is given as a function of $P_1$ and the higher-order photon probability $P_{2+}$. (d) provides a cut of (c) at $P_{2+}=0.02$. The experimental state is marked with a star.}\label{fig4} 
\end{figure}

First, we add a second dimension via \(P_{n_2}\), specifically tailored to our states of the form \(\propto \ket{0} + \ket{n_2}\). Threshold values are then calculated as linear combinations \(g_0 P_{n_2} + g_1 \mathcal{C}_{n_1, n_2}\), where \(g_0\) and \(g_1\) are optimized for each case. The results are given in Fig. \ref{fig4}. In Fig. \ref{fig4}(a), the experimental state \(\propto \ket{0} - \ket{1}\) is evaluated but fails to surpass the two-dimensional threshold. Conversely, Fig. \ref{fig4}(b) shows a set of experimental states \(\propto \ket{0} + \ket{2}\) with varying $P_2$. Some of these states successfully exceed the threshold, highlighting the dependence of coherence criteria on the additional measurement dimensions.

Finally, we introduce a third dimension for evaluating the state $\propto \ket{0}-\ket{1}$, with the higher photon probability $P_{2+}$. The threshold values are now expressed as linear combinations $g_0 P_1 + g_1\mathcal{C}_{0,1} + g_2 P_{2+}$, where the parameters $g_i$, $\alpha$, and $\xi$ are optimized. The resulting 3D plot in Fig. \ref{fig4}(c) gives the threshold as a function of the single-photon and the more-than-two-photon probabilities. The experimental state has a $P_{2+}$ value of 0.02, which corresponds to a 2D cut through Fig. \ref{fig4} (c), shown in light grey and plotted separately in Fig. \ref{fig4}(d). While the experimental state now surpasses the non-classical threshold, it still does not exceed the non-Gaussian threshold.
 
So far, we have developed a hierarchical framework to assess the non-Gaussianity of coherences using the operator $X$, which specifically targets balanced superpositions. We now extend our approach by introducing a new operator that offers a more comprehensive perspective on coherences. This operator is capable of evaluating coherences in non-perfectly balanced superpositions, enabling the analysis of any qubit state.

\vspace{0.12 cm} 
\par \textit{Qubit non-Gaussian coherence criteria.---}Qubits often exhibit unbalanced weights, \textit{i.e.}, pointing out of the equator plan of the Bloch sphere. This imbalance reduces the maximum achievable coherence. To take this into account, we define an absolute criterion for a new target state that includes an adjustable balance parameter, as $(1/\sqrt{2})(\cos(\theta) \ket{n_1} + e^{i\phi} \sin(\theta) \ket{n_2})$. The operator is modified as  
\begin{eqnarray}
S(\phi, \theta)_{n_1, n_2} &=& \sin(\theta) X_{n_1, n_2}(\phi) \nonumber \\
 &&+ \cos(\theta) [\ket{n_1}\bra{n_1}-\ket{n_2}\bra{n_2}]
\end{eqnarray}
where  and the local coherence measure is adapted to fix a specific value of $\theta$ as
\begin{eqnarray}
\mathcal{G}_{n_1, n_2}^{\theta}(\rho) = \max_{\phi} \Tr\left[S(\phi, \theta)_{n_1, n_2} \rho \right].
\end{eqnarray}
This qubit coherence $\mathcal{G}~\in~[-1, 1]$ can be interpreted as analogous to a Mach-Zehnder interferometer with a fixed phase and splitting ratio, followed by a measurement. Fixing the phase $\phi$ does not constrain the set of free states as sufficient degrees of freedom remain to align their maximum values with the chosen phase. The non-Gaussian threshold for this measure is given by 
\begin{eqnarray}
T_{Q,l}^{L,G}(\theta) = \max_{\ket{F}_{m,l}^{L,G} \in F_l^{G}} \mathcal{G}_{t}^{\theta}(\ket{F}_{m,l}^{L,G}),
\end{eqnarray}
with a similar expression for the non-classical threshold.

\begin{figure}[t!]
\includegraphics[width=0.95\columnwidth]{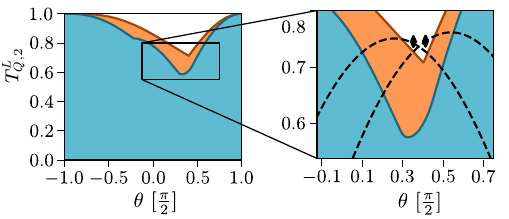}
\caption{Qubit non-classical (blue) and non-Gaussian (orange) coherence thresholds for the experimental state $\propto \ket{0}+\ket{2}$. The thresholds $T_{Q,2}^{L}$ are plotted as a function of the qubit phase~$\theta$. The zoom shows the thresholds around their respective minima. The dotted lines represent the coherence measure applied to two experimental states that show non-Gaussian qubit coherences, as marked by diamonds.}\label{fig5} 
\end{figure}

We apply the qubit-coherence threshold to the experimental state \(\propto \ket{0} + \ket{2}\), which has already shown strong performance in previous one-dimensional coherence tests but not for an absolute criterion. The threshold is now calculated for varying angles \(\theta\), with the results presented in Fig. \ref{fig5}. The threshold reaches a minimum for certifying qubit non-Gaussian coherences around \(\theta = 0.4 \times \pi/2\), which we define as the minimal requirement for a qubit to be classified as having non-Gaussian coherences.

In the zoom of Fig. \ref{fig5}, two of our states are represented by dotted lines, with diamonds indicating their best performances relative to the threshold. Both states surpass the threshold at and near the minimal requirement. This confirmes their coherence properties and validates their classification as qubit non-Gaussian states.

\vspace{0.12 cm} 
\par \textit{Conclusion.---}We have developed and experimentally tested a hierarchical, threshold-based framework for assessing non-Gaussian coherences in superpositions of Fock states. This framework introduce multiple levels of criteria, including relative thresholds that incorporate partial state information. We also extended the study to arbitrary qubit superpositions. Experimental superpositions were shown to surpass these relative thresholds, as well as the qubit coherence one. The introduced hierarchy offers a novel perspective to probe and harness non-Gaussianity. We anticipate that this framework can be further generalized to encompass more complex coherences, including those found in states such as coherent-state superpositions and Gottesman-Kitaev-Preskill states. Such developments could further expand the ability to evaluate and leverage quantum states for achieving quantum advantage in diverse settings.\\

\begin{acknowledgments}
This work was supported by the French National Research Agency via the France 2030 project Oqulus (ANR-22-PETQ-0013). L.L. acknowledges the support from the project No. 23-06015O and R.F from the project 21-13265X, both of the Czech Science Foundation. J.L. is a member of the Institut Universitaire de France. 
\end{acknowledgments}

\onecolumngrid

\clearpage
\setcounter{figure}{0}
 \renewcommand\figurename{\textbf{Supplementary Figure }}

\appendix

\section{Formal aspects of hierarchical properties}
We consider the Hilbert space $\mathcal{H}_N$ spanned by the Fock states $|n\rangle$ with $n\leq N$, and a formally ordered sequence of quantum properties $P_{k,N}$ with an index $k \in \{1,...,N\}$ that states in $\mathcal{H}_N$ exhibit. We require that $P_{k,N}$ is a property of at least one state $|\psi_k\rangle \in \mathcal{H}_N$. Due to the ordering of properties $P_{k,N}$, we expect the following:
\begin{itemize}
	\item The highest property $P_{N,N}$ exploits the state $|N\rangle$ as a resource.
	\item If a state $|\psi\rangle \in \mathcal{H}_N$ exhibits the property $P_{k,N}$, the same state exhibits the property $P_{k-1,N}$ as well.
	\item $P_{k,N}$ is a stronger property than $P_{k-1,N}$.
\end{itemize}
To formalize these demands, we associate a property $P_{k,N}$ with states $|\psi_{1,k}\rangle,..., |\psi_{m,k}\rangle$ such that $|\psi_{i,k}\rangle \in \mathcal{H}_{N}$ for all $i \in \{1,...,m\}$. We say that the states $|\psi_{i,k}\rangle$ (with $i\in \{1,...,m\}$) target the property $P_{k,N}$, \textit{i.e.}, $|\psi_{i,k}\rangle$ corresponds to the representative states of the property $P_{k,N}$. Further, we define $\mathcal{P}_{k,N}=\{|\psi_{1,k}\rangle,...,|\psi_{m,k}\rangle\}$ as a set of the target states linked to the property $P_{k,N}$. Since we require that the properties $P_{k,N}$ are ordered, we impose on $\mathcal{P}_{k,N}$ the conditions
\begin{itemize}
	\item $|\psi\rangle \in \mathcal{P}_{N,N} \Rightarrow|\psi\rangle \notin \mathcal{H}_{N-1}$
	\item $\mathcal{P}_{k,N}\subset \mathcal{P}_{k-1,N}$
	\item $\mathcal{P}_{k-1,N}\neq \mathcal{P}_{k,N}$.
\end{itemize}
These statements postulate a general concept of ordering properties associated with $\mathcal{H}_N$.

Alternatively, we specify a property $P_{k,N}$ based on introducing the Hilbert subspace $\mathcal{H}_{k} \subset \mathcal{H}_N$ that does not enable the property $P_{k,N}$. In other words, any state $|\psi\rangle \in \mathcal{H}_k$ does not exhibit the property $P_{k,N}$. In general, we allow for a set of Hilbert subspace $\boldsymbol{\mathcal{H}}_k=\{\mathcal{H}_{1,k},...,\mathcal{H}_{m,k}\}$ such that any state $|\psi\rangle \in \mathcal{H}_{i,k}$ with $\mathcal{H}_{i,k} \in \boldsymbol{\mathcal{H}}_k$  has not the property $P_{k,N}$. In summary, we identify a property $P_{k,N}$ by defining a set of target states $\mathcal{P}_{k,N}$ and by providing a set $\boldsymbol{\mathcal{H}}_{k}$ of one or more Hilbert subspace without the property $P_{k,N}$. To avoid contradiction, we require that 
\begin{equation}\label{SM:hierCond}
	|\psi\rangle \in \mathcal{P}_{k,N} \Rightarrow \forall \mathcal{H}\in \boldsymbol{\mathcal{H}}_k: |\psi\rangle \notin \mathcal{H}.
\end{equation}
On the contrary, we expect that $\exists \mathcal{H}\in \boldsymbol{\mathcal{H}}_k : |\psi\rangle \in \mathcal{H} \nRightarrow |\psi\rangle \notin \mathcal{P}_{k,N}$, \textit{i.e.}, we do not require that $\mathcal{P}_{k,N}$ corresponds to a set of all states with the property $P_k$ but it rather contains only the representative states of the property $P_k$.

So far, we have focused only on formal description related to pure states in $\mathcal{H}_N$. However, description in terms of a density matrix $\rho$ better characterizes realistically prepared states. To attach a property to $\rho$, we follow an operational approach relying on introducing measurement $M_{|\psi_{j,k}\rangle}$ for each state $|\psi_{j,k}\rangle \in \mathcal{P}_{k,N}$ that differentiates $|\psi_{j,k}\rangle$ from any state $\rho$ that reads
\begin{eqnarray}\label{SM:corRho}
	\rho=\sum_{i=1}^m p_i |\widetilde{\psi}_i\rangle \langle \widetilde{\psi}_i|,
\end{eqnarray}
where $|\widetilde{\psi}_i\rangle \in \mathcal{H}_{i,k}$ for each $i\in \{1,...,m\}$. We introduce the notation $\rho \in \boldsymbol{\mathcal{H}}_k$ meaning that $\rho$ is expressed as Eq. \ref{SM:corRho}. Thus, we certify the property $P_{k,N}$ of $\rho$ when we find at least one measurement $M_{|\psi_{j,k}\rangle}$ distinguishing the state $\rho$ from any $\widetilde{\rho}\in \boldsymbol{\mathcal{H}}_k$.

\section{Quantum non-Gaussian coherence criteria}

\subsubsection{Absolute criteria}
Here, we detail the derivation of the criteria for a hierarchy of quantum non-Gaussian coherence. We consider the ordering of the targeted states $|\psi_{k,l}\rangle=\left(|k\rangle+|k+l\rangle\right)/\sqrt{2}$ according to the length $l$ and irrespective to the index $k$. To operationally characterize the target state $|\psi_{k,l}\rangle$, we introduce the parameter $C_{k,l}(\rho)$ as:
\begin{eqnarray}\label{SM:Cl}
	\mathcal{C}_{k,l}(\rho)\equiv\frac{1}{2}\left\{\max_{\phi}\mbox{Tr}\left[\rho\left(e^{i\phi}|k\rangle\langle k+l|+e^{-i\phi}|k+l\rangle\langle k |\right)\right]-\min_{\phi}\mbox{Tr}\left[\rho\left(e^{i\phi}|k\rangle\langle k+l|+e^{-i\phi}|k+l\rangle\langle k |\right)\right]\right\}.
\end{eqnarray}
This definition implies that $\mathcal{C}_{k,l}(\rho) \in \left[0,1\right]$ for any state $\rho$. Further, we introduce states that achieve the minimal value $\mathcal{C}_{k,l}(\rho)=0$ for any $k$. To do that, we define the Hilbert subspace $\mathcal{H}_{m,l}\equiv \sum_{k=m}^{l+m-1}c_k |k\rangle$ and consider the states 
\begin{eqnarray}\label{SM:freeState}
	\rho_l=\sum_{m=0}^{\infty}p_m |\widetilde{\psi}_{m,l}\rangle\langle \widetilde{\psi}_{m,l}|,
\end{eqnarray}
where $|\widetilde{\psi}_{m,l}\rangle \in \mathcal{H}_{m,l}$. Any state $\rho_l$ reaches $\mathcal{C}_{k,l}(\rho_l)=0$ for all $k$. Conversely, only the target states $(|k\rangle+|k+l\rangle)/\sqrt{2}$ reach the maximal value $\mathcal{C}_{k,l}(\rho_l)=1$.

\begin{figure*}[t]
	\includegraphics[width=0.95\columnwidth]{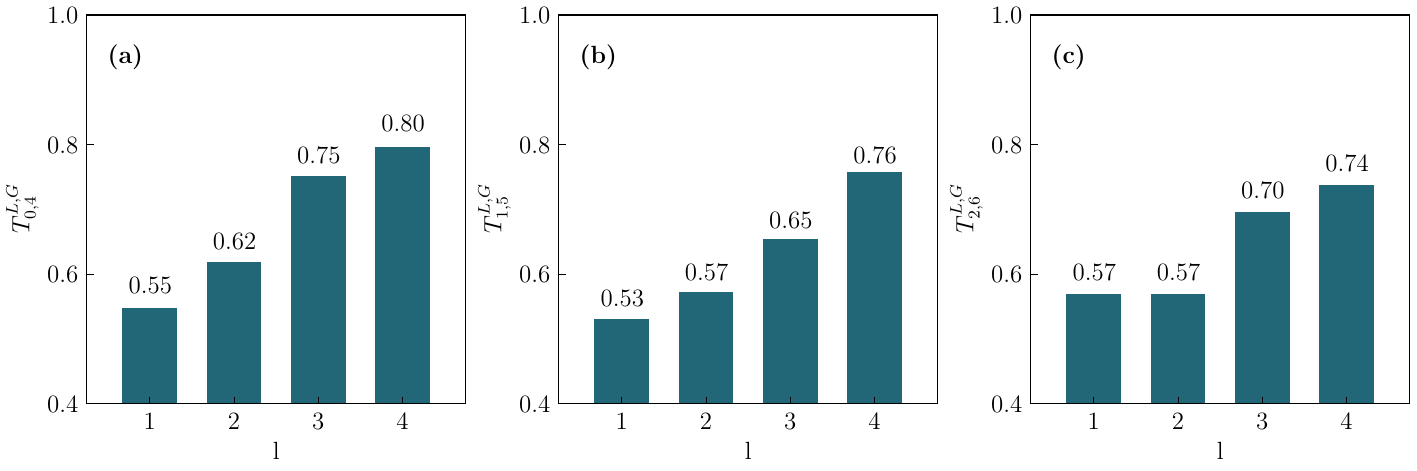}
	\caption{Examples of absolute criteria imposing conditions on experimentally achieved coherence measure $\mathcal{C}_{m,n}$. The presented criteria require that measured $\mathcal{C}_{0,4}$, $\mathcal{C}_{1,5}$ or $\mathcal{C}_{2,6}$ overcome one or more thresholds $T_{0,4}^{L,G}$ in (a), $T_{1,5}^{L,G}$ in (b) or $T_{2,6}^{L,G}$ in (c), respectively. The horizontal axis in each figure identifies the order $l$ of a tested criterion defining the set of states $S(\xi)D(\alpha)|\widetilde{\psi}_l\rangle$ with $|\widetilde{\psi}_l\rangle=\sum_{k=m}^{m+l-1}c_k |k\rangle$, which the respective threshold covers. Comparing a given coherence measure with the row of increasing thresholds enables hierarchical certification of the targeted coherence.} 
	\label{SM:fig:AbsTh}
\end{figure*}

We investigate the amount of coherence that Gaussian dynamics induces when it acts on the state $\rho_l$. The dynamics is given by:
\begin{eqnarray}
U_G(\xi,\alpha)\equiv S(\xi)D(\alpha) \quad \textrm{with} \quad S(\xi)=\exp\left[\xi \left(a^{\dagger}\right)^2-\xi^* a^2\right]\quad \textrm{and} \quad D(\alpha)=\exp\left(\alpha a^{\dagger}-\alpha^* a\right),
\end{eqnarray}
where $S(\xi)$ and $D(\alpha)$ are the squeezing and displacement operators, respectively. We need to express the overlap $g_{k,n}(\xi,\alpha)\equiv\langle k|U_G(\xi,\alpha)|n\rangle$ as a function of the complex parameters $\xi$ and $\alpha$. We introduce the phases $\phi_{\xi}$ and $\phi_{\alpha}$ defined as: $\xi=|\xi|\exp(i\phi_{\xi})$ and $\alpha=\xi \exp(i\phi_{\alpha}/2)$, which enables the parametrization of the overlap $g_{k,l}(\xi,\alpha)$ by $|\alpha|$, $|\xi|$, $\phi_{\alpha}$ and $\phi_{\xi}$. To simplify the notation, we further use in this parametrization real and positive $\alpha$ and $\xi$ instead of $|\alpha|$ and $|\xi|$. Further, we define the auxiliary function
\begin{equation}\label{SM:functionS}
	S_{k,l}(x,y,z)=\sum_{j=0}^{\min \{k,l\}}\frac{z^{k-j}H_{k-j}(x)H_{l-j}(y)}{\sqrt{k!l!(k-j)!(l-j)!}}
\end{equation}
where $H_k(x)$ is the Hermite polynomial of the order $k$. We  also introduce the parameters $a$, $b$ and $c$ as:
\begin{equation}\label{SM:abc}
	\begin{aligned}
		a&=\alpha\frac{\cosh \xi \exp(i\phi_{\alpha})+\sinh \xi \exp(-i\phi_{\alpha})}{\sqrt{\sinh 2\xi}}\\
		b&=\frac{\alpha \exp(i\phi_{\alpha})}{\sqrt{\sinh 2\xi}}\\
		c&=\frac{i}{2} \sinh \xi.
	\end{aligned}
\end{equation}
Then, $g_{k,l}(\xi,\alpha,\phi_{\alpha},\phi_{\xi})$ can read as \cite{Kral1990}:
\begin{equation}\label{SM:gkl}
	\begin{aligned}
		g_{k,l}(\xi,\alpha,\phi_{\alpha},\phi_{\xi})&=\frac{\left(\tanh \xi\right)^{n/2}\exp\left[i\phi_{\xi}(k-l)/2\right]}{\sqrt{\cosh \xi}\left(\sinh 2\xi\right)^{m/2}}\\
		&\times S_{m,n}\left(a,b,c\right)\\
		&\times \exp \left[-\frac{\alpha^2}{2}\left(1+e^{2 i\phi_{\alpha}} \tanh \xi \right)\right].
	\end{aligned}
\end{equation}
Finally, we define the matrix $\mathcal{U}_G^{(m,k,l)}$ as:
\begin{equation}\label{SM:matG}
	\begin{aligned}
		&\mathcal{U}_G^{(m,k,l)}(\xi,\alpha,\phi_{\alpha},\phi_{\xi},\phi)\equiv\\
		&\sum_{i=m}^{l+m-1}\sum_{j=m}^{l+m-1} \left[e^{i\theta} g_{i,k}(\xi,\alpha,\phi_{\alpha},\phi_{\xi})g_{j,k+l}^*(\xi,\alpha,\phi_{\alpha},\phi_{\xi})+ e^{-i\theta}g_{i,k+l}(\xi,\alpha,\phi_{\alpha},\phi_{\xi}) g_{j,k}^*(\xi,\alpha,\phi_{\alpha},\phi_{\xi})\right]|i\rangle \langle j|.
	\end{aligned}
\end{equation}
Putting Eqs.~(\ref{SM:functionS} - \ref{SM:matG}) together allows us to obtain the following parametrization
\begin{equation}\label{SM:parametrizationC}
	\mathcal{C}_{k,l}\left(|\psi_{m,l}\rangle \langle \psi_{m,l}| \right)=\max_{\phi}\langle \widetilde{\psi}_{m,l}| \mathcal{U}_G^{(m,k,l)}(\xi,\alpha,\phi_{\alpha},\phi_{\xi},\phi) |\widetilde{\psi}_{m,l}\rangle,
\end{equation}
where $|\psi_{m,l}\rangle \equiv S\left(\xi e^{i\phi_{\xi}}\right)D\left(\alpha e^{i\phi/2- i\phi_{\xi}}\right)|\widetilde{\psi}_{m,l}\rangle$ with $|\widetilde{\psi}_{m,l}\rangle \in \mathcal{H}_{m,l}$. This analytical parametrization is necessary for the effective derivation of the thresholds used to certify the quantum non-Gaussian coherence.

To reject the Gaussian dynamics acting on the states given in Eq.~(\ref{SM:freeState}), we derive the maximum value
\begin{equation}\label{SM:thres}
	T_{k,l}^{L,G}=\max_{\rho_l \in \boldsymbol{\mathcal{H}}_l,\xi>0,\alpha>0,\phi_{\alpha},\phi_{\xi},\phi} 	\mathcal{C}_{k,l}\left(U_G\rho_l U_G^{\dagger}\right),
\end{equation}
where $\boldsymbol{\mathcal{H}}_l\equiv \cup_{m=0}^{\infty} \{\mathcal{H}_{m,l}\}$ corresponds to a set of Hilbert subspaces and $G\equiv S\left(\xi e^{i\phi_{\xi}}\right)D\left(\alpha e^{i\phi/2- i\phi_{\xi}}\right)$ is Gaussian dynamics.
To derive the maximum in Eq.~(\ref{SM:thres}), we consider the triangular inequality to prove that
\begin{equation}\label{SM:triangleInEq}
	\mathcal{C}_{k,l}\left(\rho\right)\leq \sum_{m=0}^{\infty}p_m \mathcal{C}_{k,l}\left(|\psi_{m}\rangle \langle \psi_{m}| \right),
\end{equation}
where $\rho=\sum_{m=0}^{\infty}p_m |\psi_{m}\rangle \langle \psi_{m}|$. Eq.~(\ref{SM:triangleInEq}) implies that the maximum $T_{k,l}^{L,G}$ occurs for a pure state, \textit{i.e.} maximizing state $\rho_l$ always reads as $\rho_l=|\widetilde{\psi}_{m,l}\rangle \langle \widetilde{\psi}_{m,l}|$. Consequently, the threshold $T_{k,l}^{L,G}$ is a result of maximizing over Gaussian dynamics $U_G$, states $|\widetilde{\psi}_{m,l}\rangle \in \mathcal{H}_{m,l}$ and the index $m$. Maximizing over $|\widetilde{\psi}_{m,l}\rangle$ can be calculated effectively using the method in \cite{Fiurasek2022}. Based on this method, we derive the function
\begin{equation}\label{SM:maxAbsTh}
	\mathcal{C}_{k,l}(\xi,\alpha,\phi_{\alpha},\phi_{\xi},m)\equiv \max_{|\widetilde{\psi}_{m,l}\rangle\in \mathcal{H}_{m,l}} \mathcal{C}_{k,l}\left(U_G|\widetilde{\psi}_{m,l}\rangle \langle \widetilde{\psi}_{m,l}|U_G^{\dagger}\right),
\end{equation}
which depends only on the parameters $\xi$, $\alpha$, $\phi_{\alpha}$ and $\phi_{\xi}$ defining the Gaussian dynamics and the index $m$. Finally, we numerically optimize the function $\mathcal{C}_{k,l}(\xi,\alpha,\phi_{\alpha},\phi_{\xi},m)$ to achieve the thresholds $T_{k,l}^{L,G}$. Fig.~\ref{SM:fig:AbsTh} illustrates examples of such thresholds.

\begin{figure*}[t]
	\includegraphics[width=\columnwidth]{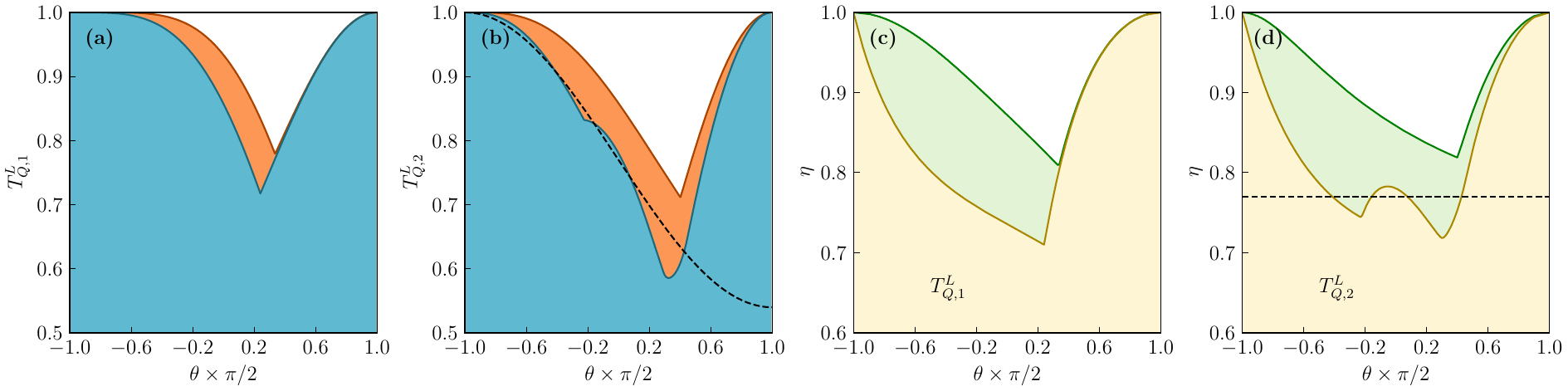}
	\caption{Qubit-coherence non-Gaussian and non-classical thresholds for the measurement of $\mathcal{G}_{0,1}^{\theta}$ in (a) and $\mathcal{G}_{0,2}^{\theta}$ in (b) with respect to $\theta$. In both figures, the orange area depicts the non-Gaussian threshold and the blue area illustrates the non-classical threshold. In (b), only the second-order $l=2$ threshold for the quantum non-Gaussian (non-classical) coherence is shown. The dashed black line in (b) shows a position of model states $\rho_{\theta,2}$ introduced in Eq.~(\ref{SM:models}) with the transmission $\eta=0.77$ that beat the nonclassical threshold for two disconnected intervals of $\theta$. For the model state $\rho_{\theta,1}$ and $\rho_{\theta,2}$ in Eq.~(\ref{SM:models}), the green (yellow) line in (c) and (d), respectively, demonstrates the threshold values of $\eta$ (characterizing the losses of the states $\rho_{\theta,1}$ and $\rho_{\theta,2}$) that enable beating the respective quantum non-Gaussian (nonclassical) threshold. The dashed black line in (d) shows the model state $\rho_{\theta,2}$ having $\eta=0.77$, which is represented by the dashed black line in (b). This state beats the threshold for $\theta \in (-0.41 \pi/2, -0.16 \pi/2)$ and $\theta \in (0.075 \pi/2,0.43 \pi/2)$.} 
	\label{SM:fig:depth}
\end{figure*}

The coherence measure $\mathcal{C}_{k,l}$ quantifies the coherence of the states $\left(|k\rangle+|k+l\rangle\right)/\sqrt{2}$, given as a balanced superposition of the Fock state $|k\rangle$ and $|k+l\rangle$. To target the general qubit states $|\psi_{\theta,k,l}\rangle=\cos (\theta/2-\pi/4) |k\rangle+\sin (\theta/2-\pi/4) |k+l\rangle$, we allow for measurement $\mathcal{G}_{k,l}^{\theta}$ introduced as
\begin{equation}
	\mathcal{G}_{k,l}^{\theta}(\rho)=\cos \theta \mathcal{C}_{k,l}(\rho)+\sin \theta \left(\langle k+l|\rho|k+l\rangle-\langle k|\rho|k\rangle\right),
\end{equation}
which corresponds to a projection on the Bloch sphere with the Fock states $|k\rangle$ and $|k+l\rangle$ at its poles, and therefore we get $\mathcal{G}_{k,l}^{\theta}(|\psi_{\theta,k,l}\rangle)=1$. To certify the quantum non-Gaussian coherence based on $\mathcal{G}_{k,l}^{\theta}$ instead of $\mathcal{C}_{k,l}$, we analogously apply the procedure to derive the thresholds
\begin{equation}\label{SM:thres}
	T_{Q,k,l}^{L,G}=\max_{\rho_l \in \boldsymbol{\mathcal{H}}_l,\xi>0,\alpha>0,\phi_{\alpha},\phi_{\xi}} 	\mathcal{G}_{k,l}^{\theta}\left(U_G\rho_l U_G^{\dagger}\right),
\end{equation}
which cover all the values of $\mathcal{G}_{k,l}^{\theta}$, that we achieve by any state $U_G\rho_l U_G^{\dagger}$. Fig.~\ref{SM:fig:depth} depicts examples of such thresholds with respect to various values of $\theta$.

For each quantum non-Gaussian criterion, we derive its nonclassical counterpart that rejects the situations when only the displacement unitary operator affects a respective set of core states. In these cases, the derivation of the nonclassical criteria is analogous and technically simpler to the derivation of quantum non-Gaussian criteria since we maximize respective formulas over the amplitude of the displacement operator instead of maximizing over general Gaussian dynamics.

\subsubsection{Relative criteria}
As shown in the main text, the absolute non-Gaussian criteria presented so far is highly demanding. To provide intermediate steps, we introduce multi-dimensional relative criteria, which leverage additional accessible information from the measured density matrix. 

To establish these criteria, we allow for the linear combination $F_{n}^{\{gc,gn\}}(\rho)=g_c\mathcal{C}_{0,n}(\rho)+g_n P_n(\rho)$, where $g_c$ and $g_n$ are free parameters and $P_n\equiv \langle n|\rho|n\rangle$ stands for the probability of $n$ photons in a state $\rho$. We derive numerically the maximum
\begin{equation}\label{SM:maxF}
	F_{n}(g_c,g_n)\equiv \max_{\rho_l \in \boldsymbol{\mathcal{H}}_l,\xi>0,\alpha>0,\phi,\phi_{\xi}} F_{n}^{\{gc,gn\}}(U_G\rho_l U_G^{\dagger})
\end{equation}
as a function of the parameters $g_c$ and $g_n$. A criterion for the quantum non-Gaussianity requires $F_n^{\{gc,gn\}}(\rho)>F_n(g_c,g_n)$ for some $g_c$ and $g_n$. Because the definition of $F_{n}^{\{gc,gn\}}$ and $F_{n}(g_c,g_n)$ imply $F_{n}^{\{gc,gn\}}(g_c,g_n)=g_c F_{n}(1,\lambda)$ with $\lambda=g_n/g_c$, we can reformulate the criterion following as \cite{Filip2011}:
\begin{equation}\label{SM:nonG2D}
	\mathcal{C}_{0,n}(\rho)>\min_{\lambda}\left[F_n(1,\lambda)-\lambda P_n(\rho)\right],
\end{equation}
which allows us to obtain a condition directly in the measured quantities $(\mathcal{C}_{0,n},P_n)$. 

We can extend this procedure straightforwardly to criteria combining $\mathcal{C}_{0,n}$, $P_n$ and $P_{e}=1-\sum_{k=0}^{n}P_k$. First, we derive a threshold
\begin{equation}
	F_n(g_c,g_n,g_e)\equiv \max_{\rho_l \in \boldsymbol{\mathcal{H}}_l,\xi>0,\alpha>0,\phi,\phi_{\xi}} F_{n}^{\{g_c,g_n,g_e\}}(U_G\rho_l U_G^{\dagger}),
\end{equation}
 where $F_{n}^{\{g_c,g_1,g_e\}}(\rho)=g_c\mathcal{C}_{0,n}(\rho)+g_n P_n+g_e P_e(\rho)$. A criterion can be read as:
 \begin{equation}\label{SM:nonG3D}
 	\mathcal{C}_{0,n}(\rho)>\min_{\lambda_n,\lambda_e}\left[F_n(1,\lambda_n,\lambda_e)-\lambda_n P_n(\rho)-\lambda_e P_e(\rho)\right],
 \end{equation}
 where $\lambda_n=g_n/g_c$ and $\lambda_e=g_e/g_c$.
 Minimizing the right-hand side of Eq.~(\ref{SM:nonG3D}) can be achieved numerically. Computing this minimum is technically more difficult compared to the minimization in Eq.~(\ref{SM:nonG2D}) as $F_n(1,\lambda_n,\lambda_e)$ is not a smooth function.  
 
\section{Model of experimental decohered states}
We consider a qubit state $|\psi_{\theta,n}\rangle=\cos(\theta/2-\pi/4)|0\rangle+e^{i\phi}\sin(\theta/2-\pi/4)|n\rangle$ which undergoes both losses and dephasing. We characterize the losses by the efficiency $\eta$ and the dephasing by a factor $\gamma \equiv \langle e^{i\phi}\rangle$, which quantifies the amount of random changes of the phase $\phi$. Then, the decohered state $|\psi_{\theta,n}\rangle$ becomes a mixed state with a density matrix that can be written as
\begin{equation}\label{SM:models}
	\begin{aligned}
		\rho_{\theta,n}&=\left[\cos^2(\theta/2-\pi/4)+(1-\eta)^n\sin^2(\theta/2-\pi/4)\right]|0\rangle \langle 0|+\eta^n \sin^2(\theta/2-\pi/4) |n\rangle \langle n|\\
		&+\eta^{n/2}\gamma \sin(\theta/2-\pi/4) \cos(\theta/2-\pi/4)\left(e^{i\phi}|0\rangle \langle n|+e^{-i\phi}|n\rangle \langle 0|\right).
	\end{aligned}
\end{equation}
We use this expression to address the robustness of the quantum (non-classical) non-Gaussian coherence against losses. Figure~\ref{SM:fig:depth} provides the minimal transmission efficiency to beat the coherence thresholds. As it can be seen, the imposed conditions on the losses depend on the angle $\theta$ that determines the measurement $\mathcal{G}_{0,n}^{\theta}$ used to certify quantum non-Gaussian coherence. An example of this dependence is shown in Figure~\ref{SM:fig:depth} (b) and (d), where the dashed line shows how a state with $\eta=0.77$ can perform against the threshold $T^L_{Q,2}$. This dependence highlights a possibility of optimizing $\theta$ for the given model states, thereby minimizing strictness of the requirement on $\eta$.

\section{Experimental quantum state engineering}
We generate two-mode squeezed vacuum using a type-II phase matched optical parametric oscillator pumped well below threshold \cite{LeJeannic2016}. The frequency-degenerate signal and idler photons at 1064~nm can be separated on a polarizing beam-splitter. Detecting $n$ idler photons with high-efficiency superconducting nanowire single-photon detectors heralds the creation of a Fock state $\ket{n}$ on the signal path. The generated state is emitted into a well-defined spatio-temporal mode, with a bandwidth of about 60 MHz. The state is characterized via high-efficiency homodyne detection and the full density matrix is reconstructed using a maximum-likelihood algorithm.

In order to create a superposition of the form $ c_0 \ket{0}+c_1 e^{i\varphi} \ket{1}$ a weak displacement is applied to the heralding mode \cite{Darras22}. The phase $\varphi$ is defined by the relative phase between the displacement and the heralding path, while the weights $c_0, c_1$ are controlled by the relative count rates of the two paths. Upon a single heralding event, the signal state is projected onto the desired superposition, with a measured single-photon heralding efficiency of $72\%$.

A different configuration, still using the same OPO, is used to create a superposition of the form $c_0 \ket{0}+c_2 e^{i\varphi} \ket{2}$. Here the two-mode squeezing is not perfectly separated into signal and idler but instead slightly mixed, typically by about $0.5 \%$. This mixing induces correlations that, upon a heralding detection of two photons, can create a coherent superposition of Fock states \cite{Huang2015}. 

Our states achieve a fidelity of $\mathcal{F}_1=82\%$ and $\mathcal{F}_2=83\%$ with the ideal qubits, placing them among the purest photonic qubits generated to date. These states are strongly non-Gaussian, as verified by their Wigner functions shown in Fig. 3 of the main text. In this work we focus on testing the non-Gaussianity of their coherences.

\section{Experimental density matrices and uncertainties}
High confidence in the measured density matrix elements is necessary for verifying non-Gaussian coherences. In this work, we use the bootstrapping or case resampling method. The only assumptions required are the size of the phase space in which we reconstruct our states, here up to Fock state $\ket{4}$, and the property that density matrices must be semidefinite positive. This approach is widely used in quantum state tomography across various platforms due to its ease of implementation and its ability to produce results comparable to other commonly employed methods, including Bayesian inference \cite{Schmied2016}.

The process is as follows: after determining a density matrix from the experimental data, we generate multiple \qu{new} datasets using the same measurement operators (\textit{i.e.}, quadrature operators). Each simulated dataset represents a possible measurement outcome instead of the actual data we collected. We then perform tomography on each dataset, yielding a set of density matrices. From these, we calculate the associated standard deviation and error bars for each coefficient in the density matrix.

For this work we generated $500$ datasets as we observed a convergence of the uncertainties starting from about $300$ datasets. Each dataset consists of $40000$ quadrature measurements, matching the number of measurements used to reconstruct a state from experimental data. We found error bars for the coherences of our states to be smaller than $0.01$, making them invisible given the size of the markers representing the data points in this paper. 

For the sake of completeness, we provide the density matrices used in the Fig. 3 of the main text, with the associated uncertainties on the coefficients. 
\begin{itemize}
\item{For the state $\propto \lvert 0 \rangle - \lvert 1 \rangle$, the real part of the density matrix is

\[
\left[
\begin{array}{M M M M M}
    0.619 & 0.004 & -0.325 & 0.002 & -0.0125 & 0.001 & 0.001 & 0.001 & 0.007 & 0.001 \\
    -0.325 & 0.002 & 0.359 & 0.005 & -0.006 & 0 & 0.006 & 0 & 0.001 & 0.001 \\
    -0.0125 & 0.001 & -0.006 & 0 & 0.01 & 0.003 & -0.003 & 0.001 & 0.005 & 0 \\
    0.001 & 0.001 & 0.006 & 0 & -0.003 & 0.001 & 0.003 & 0.001 & -0.002 & 0.001 \\
    0.007 & 0.001 & 0.001 & 0 & 0.005 & 0 & -0.002 & 0.001 & 0.009 & 0.001
\end{array}
\right]
\]}

The imaginary part is given by

\[
\left[
\begin{array}{M M M M M}
     0 & 0 & -0.001 & 0.001 &  0.006 & 0.002 &  0.006 & 0.002 & -0.003 & 0 \\
  0.001 & 0.001 &     0 & 0 &  0.006 & 0.001 & -0.002 & 0.002 & -0.008 & 0 \\
 -0.006 & 0.002 & -0.006 & 0.001 &     0 & 0 &     0 & 0.001 & -0.003 & 0.001 \\
 -0.006 & 0.001 &  0.002 & 0.001 &     0 & 0.001 &     0 & 0 &      0 & 0.001 \\
 -0.007 & 0 &      0.001 & 0 &      0.003 & 0.001 &     0 & 0.001 &     0 & 0
\end{array}
\right]
\]

\item{For the $\propto \lvert 0 \rangle + \lvert 2 \rangle$, the real part of the density matrix is

\[
\left[
\begin{array}{M M M M M}
     0.526 & 0.004 &  -0.007 & 0.002 &  0.348 & 0.002 & -0.001 & 0.002 &  -0.011 & 0.001 \\
    -0.007 & 0.002 &  0.045 & 0.005 &  0.003 & 0.001 &  0.019 & 0.002 & -0.003 & 0.001 \\
     0.348 & 0.002 &  0.003 & 0.001 &  0.411 & 0.005 &  0 & 0.001 & -0.010 & 0.001 \\
    -0.001 & 0.002 &  0.019 & 0.002 &  0 & 0.001 &  0.015 & 0.004 &    -0.002 & 0.001 \\
    -0.011 & 0.001 & -0.003 & 0.001 &  -0.010 & 0.001 & -0.002 & 0.001 & 0.004 &  0.001
\end{array}
\right]
\]

The imaginary part is given by

\[
\left[
\begin{array}{M M M M M}
     0 & 0 &  -0.010 & 0.001 & -0.105 & 0.001 &  0.005 & 0.001 &  -0.015 & 0.002 \\
 0.010 & 0.001 &     0 & 0 & 0.008 & 0.001 & -0.004 & 0.002 &  0 & 0 \\
  0.105 & 0.001 &  -0.008 & 0.001 &     0 & 0 &     0.003 & 0.002 &  -0.001 & 0 \\
  -0.005 & 0.001 &  0.004 & 0.002 & -0.003 & 0.002 &     0 & 0 & -0.001 & 0.001 \\
 0.015 & 0.002 &  0 & 0 & 0.001 & 0 &      0.001 & 0.001 &      0 & 0
\end{array}
\right]
\]}
\end{itemize}

\end{document}